\documentclass[aps,english,prd,twocolumn,superscriptaddress,nofootinbib,preprintnumbers,10pt,notitlepage]{revtex4}
\usepackage[latin1]{inputenc}
\usepackage{graphicx}
\usepackage{amsmath,amsthm,amssymb}
\usepackage{enumitem}
\usepackage{bm}

\usepackage{amsfonts}
\usepackage{dcolumn}
\def\be{\begin{equation}}
\def\ee{\end{equation}}
\def\ba{\begin{eqnarray}}
\def\ea{\end{eqnarray}}
\def\bs{\begin{subequations}}
\def\es{\end{subequations}}

\usepackage{color}

\newcommand{\beqa}{\begin{eqnarray}}
\newcommand{\eeqa}{\end{eqnarray}}

\makeatother

\usepackage{babel}
\makeatother
\begin{document}

\title{Observational Constraints on Quintessence: 
Thawing, Tracker, and Scaling models}

\author{Takeshi Chiba}
\affiliation{Department of Physics, College of Humanities and Sciences, Nihon University, Tokyo 156-8550, Japan}

\author{Antonio De Felice}
\affiliation{ThEP's CRL, NEP, The Institute for Fundamental Study,
Naresuan University, Phitsanulok 65000, Thailand}
\affiliation{Thailand Center of Excellence in Physics, Ministry of Education,
Bangkok 10400, Thailand}

\author{Shinji Tsujikawa}
\affiliation{Department of Physics, Faculty of Science, 
Tokyo University of Science, 
1-3, Kagurazaka, Shinjuku-ku, Tokyo 162-8601, Japan}

\begin{abstract}

For two types of quintessence models having thawing and 
tracking properties, there exist analytic solutions for the 
dark energy equation of state $w$ expressed in terms of 
several free parameters. 
We put observational bounds on the parameters 
in such scenarios by using the recent data of 
Supernovae type Ia (SN Ia), Cosmic Microwave 
Background (CMB), and Baryon Acoustic Oscillations (BAO). 
The observational constraints are quite different 
depending on whether or not the recent 
BAO data from BOSS are taken into account. 
With the BOSS data the upper bounds of 
today's values of $w$ ($=w_0$) in thawing models 
is very close to $-1$, whereas without this data 
the values of $w_0$ away from $-1$ can be still allowed.
The tracker equation of state $w_{(0)}$ during the 
matter era is constrained to be $w_{(0)}<-0.949$ at 
95 \%\,confidence level (CL)
even without the BOSS data, so that the tracker models 
with $w$ away from $-1$ are severely disfavored.
We also study observational constraints on scaling models
in which $w$ starts to evolve from 0 in the deep matter era
and show that the transition to the equation of state  
close to $w=-1$ needs to occur at an early 
cosmological epoch.
In the three classes of quintessence models studied in this paper, 
the past evolution of the Hubble parameters in the best-fit models 
shows only less than the 2.5 \%
difference compared to the $\Lambda$CDM.

\end{abstract}

\date{\today}


\maketitle

%%%%%%%%%%%%%%
\section{Introduction}
%%%%%%%%%%%%%%

Independent observational data such as SN Ia \cite{Riess,Constitution}, 
CMB \cite{WMAP1,WMAP7}, and BAO \cite{BAO1,Percival}
suggest that about 70\% of the energy density today
consists of dark energy responsible for cosmic acceleration.
For the constant dark energy equation of state $w$
the recent joint data analysis based on SN Ia, CMB, BAO, and the 
Hubble constant measurement shows that $w$ is constrained 
to be $w=-1.013^{+0.068}_{-0.073}$ at 68 \%\,CL \cite{Suzuki}.
If we use the time-dependent parametrization $w(a)=w_0+w_a(1-a)$, 
where $a$ is the scale factor normalized as $a=1$ today, 
the two parameters $w_0$ and $w_a$ are constrained to be 
$w_0=-1.046^{+0.179}_{-0.170}$ and 
$w_a=0.14^{+0.60}_{-0.76}$ \cite{Suzuki}.

One of the simplest candidates of dark energy is the cosmological 
constant characterized by the equation of state $w=-1$,
which is consistent with the current observational data.
However, if the cosmological constant originates 
from the vacuum energy associated with particle physics, 
there is a huge gap between the theoretical and 
observed values \cite{Weinberg}. 
Instead, alternative dark energy models 
with dynamically changing $w$ -- such as 
quintessence \cite{quinpapers,Ratra}
and k-essence \cite{kespapers} -- have been proposed
(see Refs.~\cite{reviews} for reviews).

Quintessence is described by a canonical scalar field $\phi$
with a potential $V(\phi)$.
In the framework of particle physics it is generally  
difficult to accommodate a very light scalar field 
with a mass of the order of the Hubble parameter 
$H_0 \approx 10^{-33}$~eV today \cite{Carrollqui,Kolda}.
However there have been theoretical attempts to construct 
viable quintessence models in particle physics -- especially
in supersymmetric theories \cite{particlepapers}.
For example, the Pseudo-Nambu-Goldstone-Boson (PNGB) \cite{PNGB}
or axions \cite{axions} have the potential of the form
$V(\phi)=\Lambda^4 [1 \pm \cos (\phi/f)]$ with
suppressed quantum corrections.

Caldwell and Linder \cite{Caldwell05} classified quintessence models 
into two classes, depending on the evolution of $w$.
The first class corresponds to thawing models, in which 
the field is nearly frozen by a Hubble friction 
during the early cosmological epoch and it starts to evolve once 
the field mass $m_{\phi}$ drops below the Hubble 
rate $H$. In this case the evolution of $w$ is 
characterized by the growth from $-1$.
The representative potential of this class is 
the hilltop potential such as 
$V(\phi)=\Lambda^4 [1+\cos (\phi/f)]$.

The second class consists of freezing models, in which 
the evolution of the field gradually slows down because
of the shallowness of the potential at late times.
For the inverse power-law potential 
$V(\phi)=M^{4+p} \phi^{-p}$ ($p>0$) \cite{Ratra}
there is a so-called tracker solution characterized by a nearly 
constant field equation of state $w=-2/(p+2)$ during 
the matter era \cite{Zlatev}. 
In this case the solutions with different 
initial conditions approach a common trajectory (tracker) first, 
which is followed by the decrease of $w$ toward $-1$.

In addition to tracking freezing models there is another 
sub-class of freezing models associated with scaling 
solutions \cite{CLW}. In this case the field equation of state 
scales as that of the background fluid during most of 
the matter era ($w \approx 0$).
The representative potential of this class is 
$V(\phi)=V_1 e^{-\lambda_1 \phi/M_{\rm pl}}
+V_2 e^{-\lambda_2 \phi/M_{\rm pl}}$, 
where $M_{\rm pl}$ is the reduced Planck mass, 
$\lambda_1$ and $\lambda_2$ are constants with
$\lambda_1 \gg 1$ and $\lambda_2 \lesssim 1$ \cite{BCN}.
In the early matter era the potential is approximated
as $V(\phi) \simeq V_1 e^{-\lambda_1 \phi/M_{\rm pl}}$, 
which gives rise to the scaling solution characterized
by $w=0$ with the field density parameter 
$\Omega_{\phi}=3/\lambda_1^2$.
At late times the dominance of the potential 
$V_2 e^{-\lambda_2 \phi/M_{\rm pl}}$ leads to 
the rapid decrease of $w$ relative to the tracker case 
mentioned above.
Note that the potential $V(\phi)=e^{-\lambda \phi/M_{\rm pl}}
[(\phi-B)^{\alpha}+A]$ proposed in 
Ref.~\cite{Skordis} also exhibits a similar property
to that of the double exponential potential
(see also Ref.~\cite{SahniWang}).

In this paper we place observational constraints on 
three types of quintessence models: 
(i) thawing models, (ii) tracking freezing models, and
(iii) scaling freezing models 
(see Refs.~\cite{quinconpapers} for related works).
This analysis covers most of quintessence potentials
proposed in literature.

For thawing models we employ the analytic expression 
of $w$ derived in Ref.~\cite{Dutta,Chiba09} under 
the approximation that $|w+1| \ll 1$ 
(see also Ref.~\cite{SchSen}). 
The likelihood analysis with the SN Ia and BAO data was 
carried out in Ref.~\cite{CDS} (see also Ref.~\cite{Dutta}).          
We update the analysis by using the latest SN Ia data         
(Union 2.1 dataset \cite{Suzuki}) and by adding the data 
of CMB shift parameters measured by WMAP7 \cite{WMAP7}. 
We also take into account the recent BAO data of
the BOSS experiment 
\cite{Anderson:2012sa}.\footnote{Two months after the initial 
submission of this paper, new BAO data in the Ly$\alpha$
forest appeared in the redshift range $2.1 \le z \le 3.5$ \cite{Busca}.
We do not take into account this new data in our likelihood analysis.}

Note that the observational constraints on thawing models were  
carried out by using a multi-parameter extension of the 
exponential potential \cite{Clemson} and by introducing 
a statefinder hierarchy \cite{Gupta}.
Our study based on the analytic solution of $w$ is more
convenient in that it covers any quintessence potential 
having thawing properties and that $w$ is expressed 
in terms of three parameters without the need 
of introducing more free parameters. 

For tracking freezing models one of the present authors
obtained the approximate analytic formula of $w$ 
expressed in terms of two free parameters \cite{Chiba10} 
(see also Ref.~\cite{Watson}).
The likelihood analysis based on the SN Ia and BAO data 
was performed in Ref.~\cite{Chiba10}. 
We show that adding the CMB and BOSS data further strengthens
the constraints on the tracker equation of state $w_{(0)}$.
Wang {\it et al.} \cite{WangChen} placed observational bounds 
on a number of quintessence potentials having tracker properties.
Our study based on the analytic formula of $w$ is general 
enough to cover such potentials.
Moreover we show that inclusion of the BOSS BAO data 
further strengthens the bounds on $w_{(0)}$
previously derived in the literature.

For scaling freezing models it is difficult to derive an analytic
expression of $w$, so we resort to numerical simulations 
to find a viable parameter space.

This paper is organized as follows.
In Sec.~\ref{para} we briefly review the procedure 
to derive approximate analytic expressions of $w$
in thawing and tracking freezing models.
The accuracy of those approximations is also discussed
by solving the equations of motion numerically.
For scaling freezing models we show that in some cases
it is possible to fit the evolution of $w$ by using a specific
parametrization.
In Sec.~\ref{obsersec} we first explain the method 
of our likelihood analysis based on the SN Ia, CMB, and 
BAO data and then we proceed to observational constraints
on three classes of quintessence models.
Sec.~\ref{consec} is devoted to conclusions.

%%%%%%%%%%%%%%%%%%%%%
\section{Parametrizations of quintessence}
\label{para}
%%%%%%%%%%%%%%%%%%%%%

Quintessence \cite{quinpapers} is described 
by a minimally coupled 
scalar field $\phi$ with a potential $V(\phi)$.
In addition to the field $\phi$
we take into account non-relativistic matter
with an energy density $\rho_m$.
The action in such a system is given by 
\be
S=\int d^4 x\,\sqrt{-g} \left[ \frac{M_{\rm pl}^2}{2}R
-\frac12 g^{\mu \nu} \partial_{\mu} \phi \partial_{\nu} \phi
-V(\phi) \right]+S_m\,,
\ee
where $g$ is the determinant of the metric $g_{\mu \nu}$, 
$R$ is the Ricci scalar, and $S_m$ is the action for 
non-relativistic matter.
In the flat Friedmann-Lema\^{i}tre-Robertson-Walker 
background with the scale factor $a(t)$, the dynamical 
equations of motion are
\ba
& &3H^2 M_{\rm pl}^2=\rho_{\phi}+\rho_m\,,
\label{be1} \\
& &\ddot{\phi}+3H \dot{\phi}+V_{,\phi}=0\,,
\label{be2} \\
& &\dot{\rho}_m+3H \rho_m=0\,,
\label{be3}
\ea
where $H=\dot{a}/a$ is the Hubble parameter, a dot represents
a derivative with respect to cosmic time $t$, 
$\rho_{\phi}=\dot{\phi}^2/2+V(\phi)$, 
and $V_{,\phi}=dV/d \phi$. 
The pressure of the field is given by 
$P_{\phi}=\dot{\phi}^2/2-V(\phi)$.
We introduce the equation of state $w=P_{\phi}/\rho_{\phi}$
and the density parameter $\Omega_{\phi}=\rho_{\phi}/(3H^2 M_{\rm pl}^2)$
of dark energy.

{}From Eqs.~(\ref{be1})-(\ref{be3}) we obtain the following equations 
for $w$ and $\Omega_{\phi}$ \cite{Dutta,SchSen} 
(see also Refs.~\cite{CLW,Macorra}):
\ba
w' &=& (1-w) \left[-3(1+w)+\lambda \sqrt{3 (1+w) \Omega_{\phi}} \right]\,,
\label{weq} \\
\Omega_{\phi}' &=& -3w \Omega_{\phi} (1-\Omega_{\phi})\,,
\label{Omeeq} \\
\lambda' &=& -\sqrt{3(1+w) \Omega_{\phi}} (\Gamma-1) \lambda^2\,.
\label{lameq2}
\ea
where $\lambda=-M_{\rm pl}V_{,\phi}/V$, 
$\Gamma=VV_{,\phi \phi}/V_{,\phi}^2$, and a prime represents 
a derivative with respect to $N=\ln a$. 
Depending on the field potential and the initial conditions, 
there are several different cases for the evolution 
of $w$ \cite{Caldwell05}.

In the following we shall discuss possible analytic solutions of $w$
for three different cases: (i) thawing, (ii) tracking freezing, and
(iii) scaling freezing models.
The presence of analytic solutions is useful in that 
some general properties of physical parameters can be 
extracted without studying a host of quintessence potentials
separately. Moreover, if $w$ can be analytically expressed in terms of 
the redshift $z$ with several free parameters, we do not need
to integrate the background equations of motion with arbitrary 
initial conditions. This greatly simplifies the likelihood analysis 
carried out in Sec.~\ref{obsersec}.

\subsection{Thawing models}

For thawing models of quintessence the field $\phi$
is nearly frozen in the early matter era because of the 
Hubble friction, so that $w$ is close to $-1$.
One can regard $w=-1$ as the fixed point of Eq.~(\ref{weq}).
For $\lambda \neq 0$ such a point is not stable, and hence
$w$ starts to grow at the late cosmological epoch.

If we assume that $\lambda$ is nearly constant, 
one can express $w$ in terms of $\Omega_{\phi}$
by using Eqs.~(\ref{weq}) and (\ref{Omeeq}) under the 
approximation $|1+w| \ll 1$ \cite{SchSen}.
This neglects the effect of the field mass squared
$V_{,\phi \phi}$, but it is possible to derive a more 
elaborate form of $w$ with the mass term taken 
into account \cite{Dutta,Chiba09}. 
In doing so, the potential is expanded
around the initial field value $\phi_i$ up to second order, i.e.,
$V(\phi)=\sum_{n=0}^2 V^{(n)}(\phi_i)\,(\phi-\phi_i)^n/n!$.

Provided that $|w+1| \ll 1$, the evolution of the scale
factor can be approximated by that of the $\Lambda$CDM 
model, i.e., $a(t)=\left[ (1-\Omega_{\phi 0})/
\Omega_{\phi 0} \right]^{1/3} \sinh^{2/3} (t/t_{\Lambda})$, 
where $t_{\Lambda}=2M_{\rm pl}/\sqrt{3V(\phi_i)}$ and 
$\Omega_{\phi 0}$ is the today's density parameter 
of quintessence.
On using this solution, Eq.~(\ref{be2}) is integrated to give
the field in terms of the function of $t$ and hence the field equation 
of state $w(t) \simeq -1+\dot{\phi^2}/V(\phi_i)$ is known.
This process leads to the following analytic expression 
of $w$ \cite{Dutta,Chiba09}:
\begin{widetext} 
\be
w(a)=-1+(1+w_0) a^{3(K-1)} \left[
\frac{(K-F(a))(F(a)+1)^K+(K+F(a))(F(a)-1)^K}
{(K-\Omega_{\phi 0}^{-1/2})(\Omega_{\phi 0}^{-1/2}+1)^K
+(K+\Omega_{\phi 0}^{-1/2})(\Omega_{\phi 0}^{-1/2}-1)^K}
\right]^2\,,
\label{wtha}
\ee
\end{widetext} 
where $w_0$ is the value of $w$ today, and 
\ba
K &=& \sqrt{1-\frac{4M_{\rm pl}^2 V_{,\phi \phi} (\phi_i)}
{3V(\phi_i)}}\,,\label{Kdef} \\
F(a) &=& \sqrt{1+(\Omega_{\phi 0}^{-1}-1)a^{-3}}\,.
\ea
The solution (\ref{wtha}) is valid for $K^2>0$.
The equation of state (\ref{wtha}) is expressed in terms of 
the three parameters $w_0$, $\Omega_{\phi 0}$, 
and $K$.

\begin{figure}
\includegraphics[height=3.2in,width=3.4in]{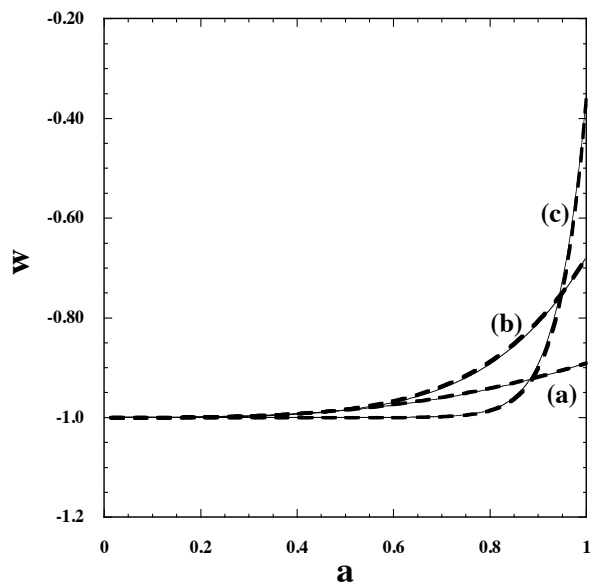}
\caption{\label{wthaw1}
The quintessence equation of state $w$ versus
$a$ for the potential (\ref{pngbpo}) with 
(a) $f/M_{\rm pl}=0.5$, $\phi_i/f=0.5$ ($K=1.9$),
(b) $f/M_{\rm pl}=0.3$, $\phi_i/f=0.25$ ($K=2.9$), and 
(c) $f/M_{\rm pl}=0.1$, $\phi_i/f=7.6 \times 10^{-4}$ 
($K=8.2$). These cases correspond to
$V_{,\phi \phi} (\phi_i)<0$, so that $K>1$.
The solid curves show numerical solutions, 
whereas the bald dashed curves describe the results derived 
from the parametrization (\ref{wtha}) with $\Omega_{\phi 0}=0.73$.}
\end{figure}

As a concrete example, 
let us consider the hilltop potential
\ba
V(\phi)=\Lambda^4 \left[ 1+\cos(\phi/f) \right]\,.
\label{pngbpo}
\ea
In this case the parameter (\ref{Kdef}) is given by 
\be
K=\left[ 1+\frac43 \left( \frac{M_{\rm pl}}{f} \right)^2
\frac{\cos (\phi_i/f)}{1+\cos(\phi_i/f)} \right]^{1/2}\,.
\ee
If $0<\phi_i/f<\pi /2$ and $\pi/2<\phi_i/f<\pi$, 
one has $K>1$ and $K<1$ respectively.
In the former case the potential is approximately 
given by $V(\phi) \approx 2\Lambda^4 [1-\phi^2/(4f^2)]$
around $\phi=0$, 
whereas in the latter case it is approximated 
as $V(\phi) \approx \Lambda^4 (\phi-\pi f)^2/(2f^2)$
around $\phi=\pi f$.

In Fig.~\ref{wthaw1} we plot the numerical evolution of $w$
versus $a$ for $K>1$ with several different values of 
$f$ and $\phi_i$.
The bald dashed curves correspond to the results derived
by the analytic expression (\ref{wtha}), which show good 
agreement with the numerically integrated solutions for 
$w_0 \lesssim -0.3$.
For $K$ larger than 10, the initial displacement of the field  
is required to be close to $0$ to avoid its rapid roll 
down. In such cases the field mass is largely 
negative, which leads to the tachyonic instability of 
field perturbations.
If the field reaches the potential minimum by today and 
it starts to oscillate, numerical simulations show 
that Eq.~(\ref{wtha}) is no longer reliable.
We set the prior $K<10$ in the likelihood analysis 
of Sec.~\ref{thawconsec}.

The analytic estimation (\ref{wtha}) starts to lose 
accuracy for $K$ smaller than 1.
This reflects the fact that the field is initially 
located away from the potential maximum.
Then the Taylor expansion around 
$\phi=\phi_i$ tends to be more inaccurate
because of the rapid variation of the field.
Numerically we find that the analytic solution
(\ref{wtha}) is reliable for $0.5 \lesssim K<1$
and $w_0 \lesssim -0.8$.

\subsection{Tracking freezing models}

In Eq.~(\ref{weq}) there is another fixed point given by 
\be
\Omega_{\phi}=\frac{3(1+w)}{\lambda^2}\,,
\label{Ometracker}
\ee
along which $w$ is constant.
This corresponds to the tracker that attracts the solutions 
with different initial conditions to a common trajectory.
The condition under which the tracking occurs 
is \cite{Zlatev}\footnote{A similar condition 
for the k-essence Lagrangian $P(\phi,X)=V(\phi)W(X)$ with 
$X=\dot\phi^2/2$ is given by $\Gamma>3/2$ \cite{chiba2002}. } 
\be
\Gamma>1\,.
\ee
In this case the variable $\lambda$ approaches 0.

{}From (\ref{Ometracker}) it follows that 
$\Omega_{\phi}'/\Omega_{\phi}=-2\lambda'/\lambda$.
Using this relation with Eqs.~(\ref{Omeeq}) and (\ref{lameq2})
in the regime $\Omega_{\phi} \ll 1$, 
the field equation of state along the tracker is given by 
\be
w=w_{(0)} \equiv -\frac{2(\Gamma-1)}{2\Gamma-1}\,.
\label{0th}
\ee
For the potential $V(\phi)=M^{4+p}\phi^{-p}$ ($p>0$) one has 
$\Gamma=1+1/p$ and hence $w_{(0)}=-2/(p+2)$
\cite{Zlatev,Watson}.

The result (\ref{0th}) was derived by neglecting the contribution of 
$\Omega_{\phi}$, but its effect can be accommodated
by dealing with $\Omega_{\phi}$ as a perturbation to the 
0-th order solution (\ref{0th}) \cite{Chiba10}.
We consider the first-order perturbation $\delta w$ around $w_{(0)}$
and then approximate $\Omega_{\phi}(a)$ by the 
0-th order solution 
\be
\Omega_{\phi}(a)=\frac{\Omega_{\phi 0}a^{-3w_{(0)}}}
{\Omega_{\phi 0}a^{-3w_{(0)}}+1-\Omega_{\phi 0}}\,.
\label{Omepa}
\ee
For the models in which $\Gamma$ is nearly constant, 
we obtain the following analytic solution \cite{Chiba10}: 
\begin{widetext} 
\ba
w(a) &=& w_{(0)}+ \sum_{n=1}^{\infty} 
\frac{(-1)^{n-1} w_{(0)} (1-w_{(0)}^2)}
{1-(n+1)w_{(0)}+2n(n+1) w_{(0)}^2}
\left( \frac{\Omega_{\phi}(a)}{1-\Omega_{\phi} (a)}
\right)^{n} \\
&=& w_{(0)}+\frac{(1-w_{(0)}^2)w_{(0)}}{1-2w_{(0)}+4w_{(0)}^2}
\Omega_{\phi} (a)+\frac{(1-w_{(0)}^2)w_{(0)}^2 (8w_{(0)}-1)}
{(1-2w_{(0)}+4w_{(0)}^2)(1-3w_{(0)}+12w_{(0)}^2)} 
\Omega_{\phi}(a)^2 \nonumber \\
& &+\frac{2(1-w_{(0)}^2)w_{(0)}^3 (4w_{(0)}-1)(18w_{(0)}+1)}
{(1-2w_{(0)}+4w_{(0)}^2)(1-3w_{(0)}+12w_{(0)}^2)
(1-4w_{(0)}+24w_{(0)}^2)} \Omega_{\phi}(a)^3+\cdots,
\label{wtrack}
\ea
\end{widetext} 
where in the second and third lines we carried out the expansion 
$(\Omega_{\phi}(a)/(1-\Omega_{\phi}(a)))^n
=\Omega_{\phi}(a)^n (\sum_{m=0}^{\infty} \Omega_{\phi}(a)^m)^n$.
The equation of state (\ref{wtrack}) is expressed in terms of 
the two parameters $w_{(0)}$ and $\Omega_{\phi 0}$.
Numerically we confirm that the approximated formula (\ref{wtrack}) 
tends to approach the full numerical solution by adding higher-order terms 
of $\Omega_{\phi} (a)$.

\subsection{Scaling freezing models}
\label{scalingfresec}

The scaling solution \cite{CLW} is a special case of a tracker along which 
$\Omega_{\phi}$ in Eq.~(\ref{Ometracker}) is constant
with $0<\Omega_{\phi}<1$.
{}From Eq.~(\ref{Omeeq}) it then follows that $w=0$
and hence $\Omega_{\phi}=3/\lambda^2$ during the 
matter era. Since $\lambda$ is constant, one has
$\Gamma=1$ from Eq.~(\ref{lameq2}).
This corresponds to the exponential potential 
$V(\phi)=V_0 e^{-\lambda \phi/M_{\rm pl}}$, where
$V_0$ is a constant\footnote{For the k-essence Lagrangian $P(\phi, X)$ 
(where $X=-g^{\mu \nu} \partial_{\mu} \phi \partial_{\nu} \phi/2$)
the condition for the existence of scaling solutions
restricts the Lagrangian in the form $P=X\,g (Y)$, where 
$g$ is an arbitrary function in terms of 
$Y=Xe^{\lambda \phi/M_{\rm pl}}$ \cite{kesscaling}.
Quintessence with the exponential potential corresponds to 
the choice $g(Y)=1-V_0/Y$. }.
In this case, however, the field equation of state
is the same as that of the background fluid ($w=0$), 
so that the system does not enter the phase of
cosmic acceleration.

This problem can be circumvented for the 
following model \cite{BCN}
\be
V(\phi)=V_1 e^{-\lambda_1 \phi/M_{\rm pl}}+
V_2 e^{-\lambda_2 \phi/M_{\rm pl}}\,,
\label{doublepo}
\ee
where $\lambda_i$ and $V_i$ ($i=1, 2$) are constants.
If $\lambda_1 \gg 1$ and $\lambda_2 \lesssim 1$
then the solution first enters the scaling regime 
characterized by $\Omega_{\phi}=3(1+w_m)/\lambda_1^2$, 
where $w_m$ is the equation of state of the background 
fluid. {}From the bound coming from big bang 
nucleosynthesis \cite{Bean} 
there is a constraint $\Omega_{\phi}<0.045$ (95\,\% CL) 
during the radiation era ($w_m=1/3$), 
which translates into the condition $\lambda_1>9.4$.
The scaling matter era ($\Omega_{\phi}=3/\lambda_1^2$, $w=0$)
is followed by the dark energy dominated epoch
driven by the presence of the potential 
$V_2 e^{-\lambda_2 \phi/M_{\rm pl}}$.
If $\lambda_2^2<3$, the solution approaches another attractor 
characterized by $\Omega_{\phi}=1$ and 
$w=-1+\lambda_2^2/3$ \cite{CLW}.
The cosmic acceleration occurs for $\lambda_2^2<2$.

\begin{figure}
\includegraphics[height=3.2in,width=3.4in]{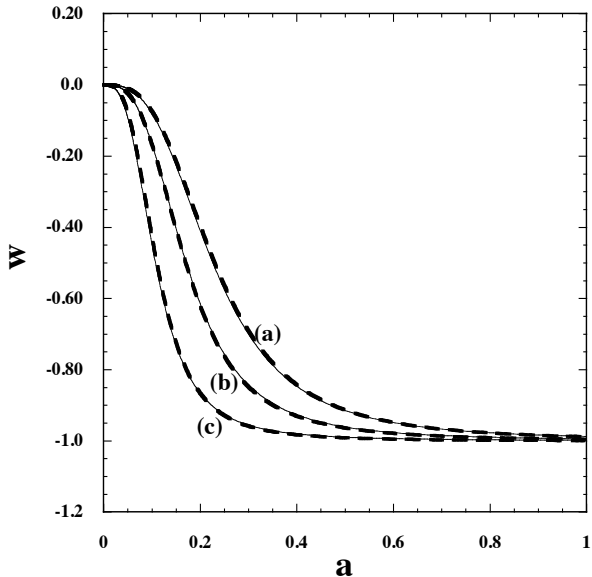}
\caption{\label{wscaling}
The quintessence equation of state $w$ versus
$a$ for the potential (\ref{doublepo}) with 
(a) $\lambda_1=10$, $\lambda_2=0$, 
(b) $\lambda_1=15$, $\lambda_2=0$, and 
(c) $\lambda_1=30$, $\lambda_2=0$.
The solid curves show the numerical solutions, 
whereas the dashed curves represent
the results derived from the parametrization 
(\ref{LHpara}) with $w_p=0$ and $w_f=-1$.
Each dashed curve corresponds to 
(a) $a_t=0.23$, $\tau=0.33$, 
(b) $a_t=0.17$, $\tau=0.33$, and 
(c) $a_t=0.11$, $\tau=0.32$.
}
\end{figure}

The onset of the transition from the scaling matter era to 
the dark energy dominated epoch depends on 
the parameters $\lambda_1$, $\lambda_2$, and 
$V_2/V_1$.
Numerically we find that the transition redshift is not very 
sensitive to the choice of $V_2/V_1$, so we study
the case $V_2=V_1$.
In Fig.~\ref{wscaling} we plot the numerical evolution of $w$
for $\lambda_2=0$ with three different values of $\lambda_1$.
For larger $\lambda_1$ the transition to the dark energy 
dominated epoch occurs earlier.

It is possible to accommodate the above variation of $w$
analytically by using the parametrization proposed by 
Linder and Huterer \cite{Linder05} 
(see also Refs.~\cite{Bassett}): 
\be
w(a)=w_f+\frac{w_p-w_f}{1+(a/a_t)^{1/\tau}}\,,
\label{LHpara}
\ee
where $w_p$ and $w_f$ are asymptotic values of $w$
in the past and future respectively, $a_t$ is the scale
factor at the transition, and $\tau$ describes the width 
of the transition.
The scaling solution during the matter era 
corresponds to $w_p=0$.
For $\lambda_2=0$ one has $w_f=-1$, 
so that the parametrization (\ref{LHpara}) reduces to 
$w(a)=-1+[1+(a/a_t)^{1/\tau}]^{-1}$.
Figure \ref{wscaling} shows that the parametrization 
(\ref{LHpara}) can fit the numerical evolution of $w$
very well for appropriate choices of 
$a_t$ and $\tau$.
For $\lambda_2=0$ the transition width is around 
$\tau \approx 0.33$, while $a_t$ 
depends on the values of $\lambda_1$.
In this case one can carry out the likelihood analysis
by fixing $\tau=0.33$ and find the constraints 
on $a_t$ (see Ref.~\cite{DNT12} for a related work).

\begin{figure}
\includegraphics[height=3.2in,width=3.4in]{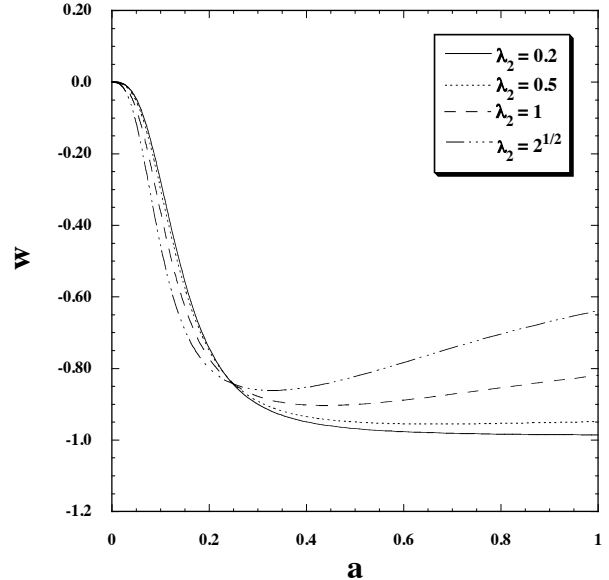}
\caption{\label{wscaling2}
The quintessence equation of state $w$ versus 
$a$ for the potential (\ref{doublepo}) with 
$\lambda_1=20$ and 
$\lambda_2=0.2, 0.5, 1, \sqrt{2}$.}
\end{figure}

If $\lambda_2 \neq 0$, then the field equation of state
finally approaches the value $w=-1+\lambda_2^2/3$.
Numerically we find that $w$ tends to have a minimum
for larger $\lambda_2$ before the solutions reach
the attractor. If $\lambda_1=20$, for example, the minimum 
appears for $\lambda_2 \gtrsim 0.3$ (see Fig.~\ref{wscaling2}).
In order to place observational bounds on $\lambda_2$
in such cases, we need to resort to numerical simulations
without using the parametrization (\ref{LHpara}).

%%%%%%%%%%%%%%%%%%%%%%%%%%%%%%%%%%%
\section{Observational constraints}
\label{obsersec}
%%%%%%%%%%%%%%%%%%%%%%%%%%%%%%%%%%%

In this section we place observational constraints on the three types
of quintessence models separately.  We use the recent SN Ia data
called Union 2.1 dataset \cite{Suzuki}, shift parameters provided by
WMAP7 \cite{WMAP7}, and the BAO distance measured by SDSS7
\cite{Percival} and by BOSS \cite{Anderson:2012sa}. 
In order to make the analysis simpler, we fix, for all the models under
consideration, the today's radiation density parameter 
to be equal to the one of the $\Lambda$CDM model.

In SN Ia observations the luminosity distance $d_L(z)=(1+z)
\int_0^z H^{-1}(\tilde{z}) d \tilde{z}$ is measured by the difference 
(distance modulus) of the apparent magnitude $m(z)$ 
and the absolute magnitude $M$, as
\be
\mu (z) \equiv m(z)-M=5 \log_{10} \left[ d_L(z)/10~{\rm pc} \right]\,.
\ee
For the observed distance modulus $\mu_{\rm obs} (z_i)$
with the errors $\sigma_{\mu, i}$, the chi square of 
the SN Ia measurement is given by 
\be
\chi_{{\rm SN\, Ia}}^{2}=\sum_{i}\frac{[\mu_{{\rm obs}}(z_{i})
-\mu_{{\rm th}}(z_{i})]^2}{\sigma_{\mu,i}^{2}}\,,
\ee
where $\mu_{{\rm th}} (z_i)$ is the theoretical value of 
$\mu (z_i)$ known for a given dark energy model.

The position of the CMB acoustic peaks is 
determined by the following parameter \cite{Bond,Koso,Muk}
\be
l_a=\frac{\pi d_a^{(c)} (z_*)}{r_s(z_*)}\,, 
\ee
where $z_*$ is the redshift at the decoupling epoch,
$d_a^{(c)} (z_*)={\cal R}/(H_0 \sqrt{\Omega_{m0}})$ 
is the comoving angular diameter distance
to the last scattering surface ($\Omega_{m0}$ is 
the matter density parameter today), and 
\be
{\cal R}=\sqrt{\Omega_{m0}} \int_0^{z_*}
\frac{dz}{H(z)/H_0}\,.
\ee
The sound horizon $r_s(z_*)$
is defined by 
\be
r_{s}(z_{*})=\int_{z_{*}}^{\infty}\frac{dz}{H(z)\,
\sqrt{3\{1+3\Omega_{b0}/
[4\Omega_{\gamma 0}(1+z)]\}}}\,.
\ee
Here $\Omega_{b0}$ and $\Omega_{\gamma 0}$
are the today's density parameters of baryons and photons,
respectively. For the redshift $z_*$ we use the fitting 
formula of Hu and Sugiyama \cite{Sugi}.\footnote{
We note that, for fixed $l_{a}$, two parameters 
${\cal R}$ and $z_*$ depend primarily on 
${\Omega_{m0}}$ and $\Omega_{b0}$, respectively. 
While $l_a$ characterizes the position of the 
CMB acoustic peaks, the parameters ${\Omega_{m0}}$ 
and $\Omega_{b0}$ are mostly related to the amplitudes of 
the peaks \cite{Sugi}.
It is possible to employ the parameter sets ($l_a,\Omega_{m0},\Omega_{b0}$)
in the likelihood analysis (as in Ref. [44]), but
we use the parameter sets ($l_a,{\cal R},z_*$)
because those are the parameters that the WMAP team 
provides an approximate covariance matrix for.}
The chi square associated with the WMAP7 
measurement is
\be
\chi_{{\rm CMB}}^{2}={\bm X}_{\rm CMB}^T
\bm{C}_{{\rm CMB}}^{-1}{\bm X}_{\rm CMB}\,,
\ee
where
${\bm X}_{\rm CMB}^T=(l_{a}-302.09,{\cal R}-1.725,z_{*}-1091.3)$, 
and the covariance matrix is given by \cite{WMAP7}
\be
\bm{C}_{{\rm CMB}}=\left(\begin{array}{ccc}
0.58269& 0.00274801& 0.318613\\
0.00274801& 0.000338358& 0.0122901\\
0.318613& 0.0122901& 0.824753
\end{array}\right)\, ,
\ee
and the inverse covariance matrix is 
\be
\bm{C}_{{\rm CMB}}^{-1}=\left(\begin{array}{ccc}
2.305 & 29.698 & -1.333\\
29.698 & 6825.27 & -113.18\\
-1.333 & -113.18 & 3.414
\end{array}\right)\,.
\ee

In BAO observations the ratio $r_{\rm BAO} (z) \equiv r_s (z_d)/D_V(z)$
is measured, where $r_s(z_d)$ is the sound horizon at which the baryons 
are released from the Compton drag of photons and
$D_V(z)$ is the effective BAO distance defined by
$D_V(z) \equiv [(\int_0^z H^{-1} (\tilde{z})d \tilde{z})^2 z/H(z)]^{1/3}$ \cite{BAO1}.
For the redshift $z_d$ we use the fitting formula of 
Eisenstein and Hu \cite{EisenHu}.
The chi square of the SDSS7 measurement is given by
\be
\chi_{{\rm BAO,SDSS7}}^{2}={\bm X}_{\rm BAO}^T
\bm{C}_{{\rm BAO}}^{-1}{\bm X}_{\rm BAO}\,,
\ee
where ${\bm X}_{\rm BAO}^T=(r_{\rm BAO}(0.2)-0.1905,
r_{\rm BAO}(0.35)-0.1097)$. 
The covariance matrix is given by \cite{Percival}
\be
\bm{C}_{{\rm BAO}}=\left(\begin{array}{cc}
3.7436\times 10^{-5}& 7.4148\times 10^{-6}\\
7.4148\times 10^{-6} & 1.2966\times 10^{-5}
\end{array}\right)\,,
\ee
and the inverse covariance matrix is
\be
\bm{C}_{{\rm BAO}}^{-1}=\left(\begin{array}{cc}
30124 & -17227\\
-17227 & 86977
\end{array}\right)\,.
\ee
We also use the BAO data from the WiggleZ and
6dFGS surveys, for which
$A_{\rm WiggleZ}(z = 0.6) = 0.452 \pm 0.018$ \cite{Blake:2011wn},
and $A_{\rm 6dFGS}(z = 0.106) = 0.526 \pm 0.028$ \cite{Beutler:2011hx}, where
$A(z)$ is defined as $A_{\rm th}(z)=D_V(z)\sqrt{\Omega_{m0} H_0^2}/z$.

Finally, we use the latest and most precise data point from 
the BOSS experiment \cite{Anderson:2012sa}, for which
\be
1/r_{\rm BAO}(z=0.57) = 13.67 \pm 0.22\,.
\label{bossdata}
\ee
Note that the error bar of this data is less than 1.7 \%.
This is the most precise distance measurement ever 
constrained from a galaxy survey. 
Moreover, as we will see below, this data puts a severe upper 
bound of $w$ close to $-1$ at $z=0.57$. 

The total chi-square from the three datasets is
\be
\chi^2=\chi_{\rm SN\,Ia}^2+
\chi_{{\rm CMB}}^{2}+\chi_{{\rm BAO}}^{2}\,,
\label{chis}
\ee
where the best-fit corresponds to the lowest value 
of $\chi^2$.

\subsection{Thawing models}
\label{thawconsec}

Let us study observational constraints on thawing models 
given by the equation of state (\ref{wtha}).
Since $K$ is weakly constrained in the three-parameter analysis, 
we first study this model by fixing the value of $K$ and vary the two parameters 
$w_0$ and $\Omega_{\phi0}$. This is the approach taken by 
Dutta and Scherrer \cite{Dutta}. However, because of a mild 
dependence (i.e.\ degeneracy) of the $\chi^2$ as a function of 
the parameter $K$, we will also perform the data analysis by 
marginalizing the $\chi^2$ over $K$ itself. 
This last procedure allows us, by hiding the information 
of $K$, to understand the measured parameters $\Omega_{\phi 0}$
and $w_0$ easily.

Although quintessence corresponds to the case $w_0>-1$, 
we also extend to the regime $w_0<-1$ in the likelihood analysis.
In fact it was shown in Ref.~\cite{DSS} that the equation of state 
of a phantom scalar field can be accommodated by the analytic 
formula (\ref{wtha}).
The likelihood results are quite different depending on whether
the recent BAO data from BOSS \cite{Anderson:2012sa} are 
included or not, so we present two constraints 
with/without the BOSS data.

Let us first focus our analysis on some particular values of $K$, 
which allow us to make comparison with the results given in \cite{Dutta}. 
Our numerical analysis gives that, in the absence of the BOSS BAO data, the 
dark energy equation of state today is constrained to be
$-1.219<w_0<-0.930$ (at 95 \%\,CL for $K=1.01$).
Dutta and Scherrer \cite{Dutta} showed that even 
the value $w_{0}=-0.7$ is allowed from the SN Ia data alone.
Adding the WMAP7 and SDSS7 data gives rise to much 
tighter bounds on $w_0$.
This is the case for $\Omega_{\phi 0}$ as well.
We obtained the bound $0.7078<\Omega_{\phi 0}<0.740$ 
(95 \%\,CL for $K=1.01$), whereas the values $0.68<\Omega_{\phi 0}<0.8$
are allowed in Ref.~\cite{Dutta}.

We also study the same case {\sl with} the latest BOSS data. 
Having so far fixed $K$, by studying the $\chi^2$ over the parameter space 
of the remaining two parameters,
we find that $w_0$ and $\Omega_{\phi 0}$ are constrained to 
be $-1.242<w_0<-0.995$ and $0.705<\Omega_{\phi 0}<0.734$, 
respectively, at 95 \%\,CL for $K=1.01$.
Rather surprisingly, the allowed parameter space in the regime
$w_0\geq -1$ is very narrow.
In particular the $\Lambda$CDM model, which corresponds to 
$w_0=-1$, is outside the $1\sigma$ observational contour.
This comes from the fact that the BOSS data (\ref{bossdata})
does not allow a large parameter space with $w>-1$ at the 
redshift $z=0.57$.
 
We have also studied the case $K=4$ and derived bounds for
$w_0$ and $\Omega_{\phi 0}$.
Without the BOSS data we obtain the bounds 
$-1.408<w_0<-0.855$ and $0.711<\Omega_{\phi 0}<0.740$ 
(95 \%\,CL, $K=4$), whereas with the BOSS data these parameters
are constrained to be $-1.460<w_0<-0.982$ and 
$0.709<\Omega_{\phi 0}<0.734$ (95 \%\,CL, $K=4$) respectively.
In the former case the upper bound of $w_0$ gets larger
than that for $K=1.01$. This reflects the fact that $w$
changes more rapidly at late times 
for larger $K$. In the presence of the BOSS data,  
however, the allowed parameter space in the regime
$w_0>-1$ is still very tiny. This same qualitative behavour 
holds for all the different values of $K$ considered in our analysis, 
as we have found that the $\chi^2$ does not vary significantly as a function of $K$.

\begin{figure}
\includegraphics[height=3.0in,width=3.2in]{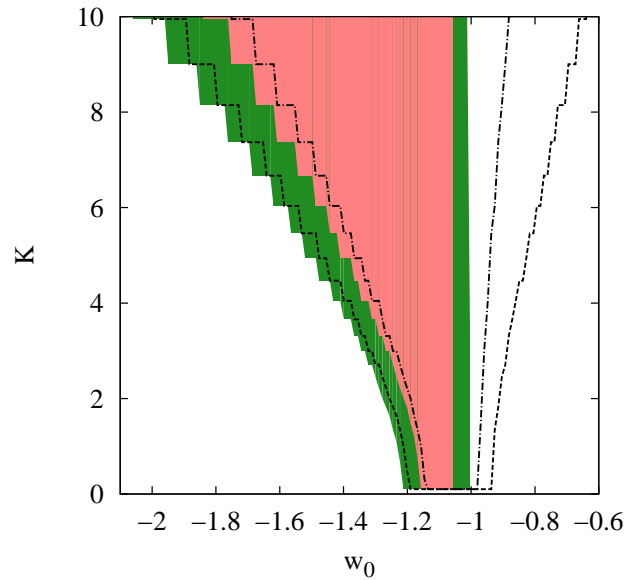}
\caption{\label{thawconK}
$1\sigma$ (red) and $2\sigma$ (green) observational contours
in the ($w_0, K$) plane marginalized over
$\Omega_{\phi 0}$. We set the prior $0.1<K<10$. 
The dot-dashed and dotted curves correspond to the $1\sigma$
and $2\sigma$ constraints without the BOSS data, respectively.}
\end{figure}

\begin{figure}
\includegraphics[height=3.0in,width=3.2in]{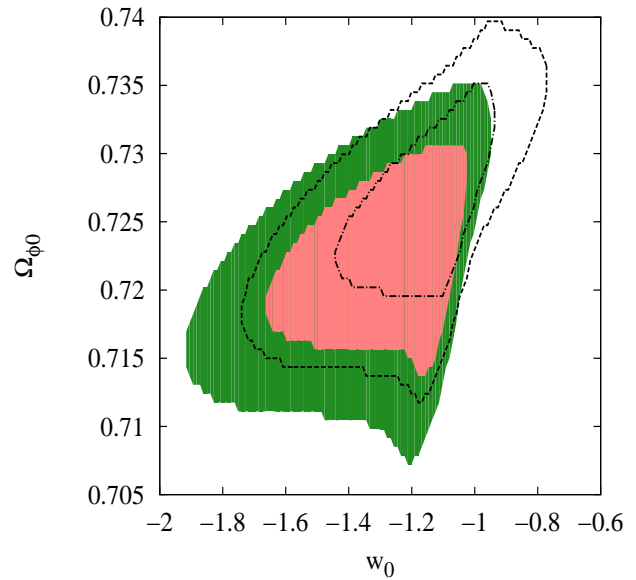}
\caption{\label{thawcon}
$1\sigma$ (red) and $2\sigma$ (green) observational contours
in the ($w_0, \Omega_{\phi 0}$) plane marginalized over
$K$. We set the prior $0.1<K<10$. 
The dot-dashed and dotted curves correspond to the $1\sigma$
and $2\sigma$ constraints without the BOSS data, respectively.}
\end{figure}

We also vary the three parameters $w_0$, 
$\Omega_{\phi 0}$, and $K$ in the likelihood analysis
with the prior $0.1<K<10$ 
updating the analysis made in Ref.~\cite{CDS}. 
We have fixed this prior for $K$ because, for $K>10$, 
the analytic expression (\ref{wtha}) is not 
completely reliable because of the rapid rolling down 
of the field along the potential with carefully 
chosen initial conditions.
With the BOSS data taken into account, the best-fit model 
parameters are found to be $w_0=-1.102$, $\Omega_{\phi0}=0.71955$, 
and $K=0.1$ with $\chi_{\rm min}^2=568.57$. 
Under the prior $w_0\geq -1$ the best-fit parameters reduce to 
those in the $\Lambda$CDM 
model as $\chi^2$ has a minimum at $w_0=-1$.

Although the case $w_0<-1$ is plagued by a ghost problem, the dynamics 
of $w$ given phenomenologically by Eq.~(\ref{wtha}) is able to fit the data quite well
for $w_0\approx-1.2$. In fact, with two parameters more than those in 
the $\Lambda$CDM, according to the Akaike Information Criterion (AIC)
(where we should add twice the number of free parameters $k$ to 
the original $\chi^2$) \cite{Akaike}\footnote{AIC assumes the infinite number 
of data points. For finite number of data points ($n$), 
AIC should be modified to the Sugiura's criterion: $\chi^2+2nk/(n-k-1)$, 
where $k$ is the number of free parameters \cite{Sugiura}.
Since in our case $n$ is large ($n>500$) and $k$ is of the order of 1, 
the difference between two criteria is very small.}, 
the best-fit corresponds to 
$\tilde{\chi}^2=568.57+2\times3=574.57$, whereas in the 
$\Lambda$CDM $\tilde{\chi}^2_{\Lambda{\rm CDM}}=573.89+2=575.89$. 
Therefore the model with $w_0<-1$, even with three parameters, 
can compete with the $\Lambda$CDM. 

In Fig.~\ref{thawconK} we show observational constraints
in the ($w_0, K$) plane marginalized over $\Omega_{\phi0}$. 
In the regime $0.1<K<1$ the constraints on $w_0$ are practically 
independent of $K$, i.e., $-1.212<w_0<-1.003$ (95 \%\,CL) with 
the BOSS data, which should be compared with the previous result 
of Ref.~\cite{Chiba10} (slightly updating the result in Ref.~\cite{CDS}), 
$-1.14<w_0<-0.92$ for $K<2$. 
In the presence of the BOSS data the allowed region shifts toward
$w$ less than $-1$, as we see in Fig.~\ref{thawconK}.

For $K>1$ the lower bound on $w$ gets smaller with increasing $K$, 
whereas the upper bound on $w$ is practically unchanged.
If $K=9.95$, for example, $w_0$ is constrained to be 
$-2.059<w_0<-1.014$ (95 \%\,CL).
For $K$ larger than the order of 1 the field equation of state can 
rapidly increase in low redshifts, but such rapid growth of $w$
is strongly disfavored from the BOSS data.
If we do not take into account the BOSS data, the growth of $w$
away from $w_0=-1$ can be still allowed.
For $K>10$ the analytic expression (\ref{wtha}) is not 
completely reliable because of the rapid rolling down 
of the field along the potential with carefully 
chosen initial conditions.

Finally, in Fig.~\ref{thawcon} we plot observational constraints 
in the ($w_0, \Omega_{\phi0}$) plane marginalized 
over $K$ with the prior $0.1<K<10$. 
Also in this case, we find the same trend already mentioned above, namely, 
in the presence of the BOSS data, the allowed region shifts toward the values of 
$w$ less than $-1$, as we see in Fig.~\ref{thawcon}.
After the marginalization over $K$
we obtain the bounds $-2.18<w_0<-0.893$ 
and $0.70265<\Omega_{\phi 0}<0.73515$ (95 \%\,CL).
If we put the prior $w_0>-1$, we find that $w_0$ is 
constrained to be $w_0<-0.849$ (68 \%\,CL) and 
$w_0<-0.695$ (95 \%\,CL).

\subsection{Tracking freezing models}

\begin{figure}
\includegraphics[height=3.0in,width=3.3in]{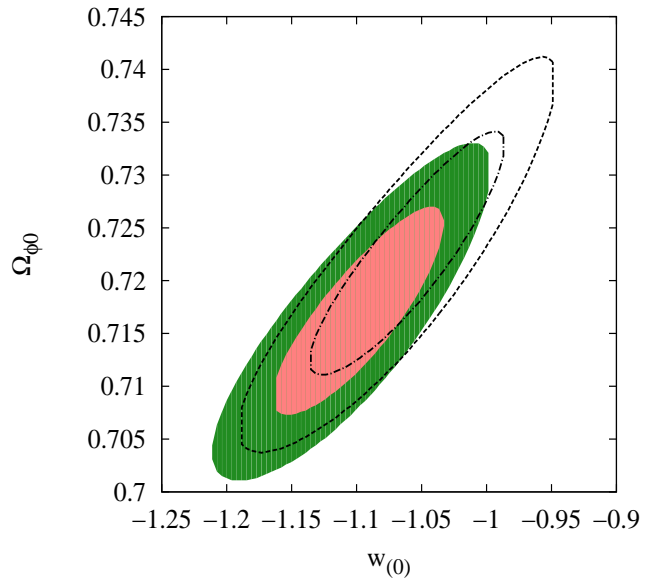}
\caption{\label{trackercon}
$1\sigma$ (red) and $2\sigma$ (green) observational contours
on tracking freezing models
in the ($w_{(0)}, \Omega_{\phi 0}$) plane.
The dot-dashed and dotted curves correspond to the $1\sigma$
and $2\sigma$ constraints without the BOSS data, respectively.}
\end{figure}

Let us proceed to observational constraints on tracker 
solutions whose equation of state is given by Eq.~(\ref{wtrack}).
Although $w_{(0)}$ is theoretically larger than $-1$ for quintessence, 
we do not put prior $w_{(0)}\geq -1$ in the actual likelihood
analysis. In Fig.~\ref{trackercon} we show $1\sigma$ and $2\sigma$
observational contours in the $(w_{(0)}, \Omega_{\phi 0})$ plane. 

Without the BAO data, the tracker equation of state 
is constrained to be $-1.188<w_{(0)}<-0.949$ (95 \% CL).
Meanwhile the analysis of Ref.~\cite{Chiba10}
based on the Constitution SN Ia and the SDSS BAO 
data gives the bound $-1.19<w_{(0)}<-0.90$ 
(95 \% CL). Hence the upper bound of $w_{(0)}$
becomes tighter by including the WMAP7 data of CMB
shift parameters.
For the potential $V(\phi)=M^{4+p}\phi^{-p}$ ($p>0$)
the $2\sigma$ constraint $w_{(0)}<-0.949$ 
translates into $p<0.107$.

If we take into account the BOSS data in the analysis, 
the best-fit model parameters are found to be $w_{(0)}=-1.097$, 
and $\Omega_{\phi 0}=0.717$ with $\chi^2_{\rm min}=568.39$.
In this case the Akaike criterion gives 
$\tilde{\chi}^2=568.39+2\times2=572.39$, 
which is smaller than the $\Lambda$CDM value 
$\tilde{\chi}^2_{\Lambda{\rm CDM}}=575.89$
with the difference more than 2.
The $2\sigma$ observational bounds are found to be 
$-1.211<w_{(0)}<-0.998$ and 
$0.701<\Omega_{\phi 0}<0.733$ (95 \% CL).
The upper bound of $w_{(0)}$ is very close 
to $-1$, which shows that the tracking quintessence
away from $-1$ is strongly disfavored from the data.

If we put the prior $w_{(0)}\geq -1$ in the analysis with the BOSS data,  
we find that the best-fit is obtained for $w_{(0)}=-1$ and 
the model coincides with the standard $\Lambda$CDM case.
We then obtain the upper bound $w_{(0)}<-0.964$ (95 \% CL).

\subsection{Scaling freezing models}
\begin{figure}
\includegraphics[height=2.9in,width=3.2in]{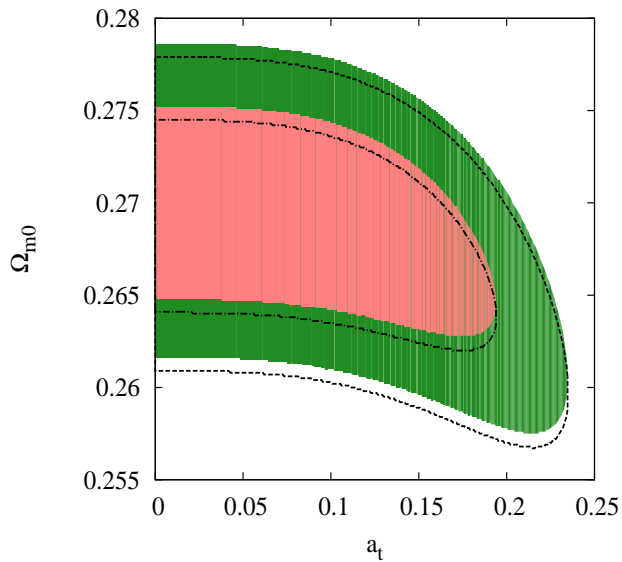}
\caption{\label{scalingcon1}
$1\sigma$ (red) and $2\sigma$ (green) observational contours
on $a_t$ and $\Omega_{m 0}$ for the parametrization (\ref{LHpara}) 
of the scaling solution with $w_p=0$, $w_f=-1$, and $\tau=0.33$.
The dot-dashed and dotted curves correspond to the $1\sigma$
and $2\sigma$ constraints without the BOSS data, respectively. }
\end{figure}
\begin{figure}
\includegraphics[height=2.9in,width=3.2in]{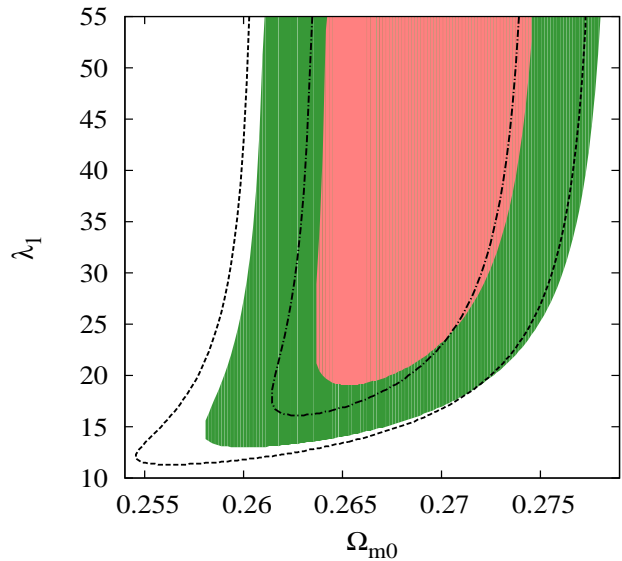}
\caption{\label{scalingcon2}
$1\sigma$ (red) and $2\sigma$ (green) observational contours
on $\lambda_1$ and $\Omega_{m 0}$ for the potential (\ref{doublepo})
with $\lambda_2=0$. 
The dot-dashed and dotted curves correspond to the $1\sigma$
and $2\sigma$ constraints without the BOSS data, respectively.}
\end{figure}

For the scaling models characterized by the potential (\ref{doublepo})
with $\lambda_2=0$ we already showed in Sec.~\ref{scalingfresec} 
that the evolution of $w$ can be well approximated 
by Eq.~(\ref{LHpara}).
This parametrization admits an exact solution of $H/H_0$ in terms of $a$ 
and the other free four parameters, which is numerically convenient 
for analyzing the data \cite{Linder05}.
In Fig.~\ref{scalingcon1} we plot observational bounds in 
the ($a_t, \Omega_{m 0}$) plane derived by using the 
parametrization (\ref{LHpara}) with 
$w_p=0, w_f=-1$, and $\tau=0.33$.
If the BOSS data are taken into account in the analysis, 
the transition redshift is constrained to be 
$a_t<0.23$ (95\,\%\,CL).
The minimum of $\chi^2$ is found to be 
$a_t=0$ with $\Omega_{m0}=0.27$, i.e.,  
the $\Lambda$CDM limit.
The case (a) shown in Fig.~\ref{wscaling} 
corresponds to the marginal one in which the model is within 
the $2\sigma$ observational contour.
This means that $w$ needs to approach $-1$ in 
an early cosmological epoch ($w<-0.8$ for the 
redshift $z<2$). Even without the BOSS data the upper bound 
on $a_t$ is practically unchanged, which reflects the fact that 
$w$ is close to $-1$ at low redshifts.

For $\lambda_2=0$ we also carry out the likelihood 
analysis without resorting to the parametrization (\ref{LHpara}).
Numerically we solve the background equations of motion
by tuning initial conditions to find the evolution which gives
the desired values of $\Omega_{m0}$ and 
$\Omega_{r0}$ (radiation density parameter)
for the potential $V(\phi)=V_1 e^{-\lambda_1 \phi/M_{\rm pl}}+V_2$.
We put the prior $\lambda_1>9.4$ coming from the constraint of
big bang nucleosynthesis.
As long as $V_1 e^{-\lambda_1 \phi/M_{\rm pl}} \gg V_2$ the 
solutions approach the scaling fixed point  
$x=y=\sqrt{6}/(2\lambda_1)$ during the matter era, 
so the initial conditions of $x$ and $y$ are irrelevant to 
the likelihood analysis.
In Fig.~\ref{scalingcon2} we show observational bounds
on the parameters $\lambda_1$ and $\Omega_{m0}$. 
The parameter $\lambda_1$ is constrained to be 
$\lambda_1>13$ (95\,\%\,CL), which is consistent 
with the results presented in Fig.~\ref{scalingcon1}.
For larger $\lambda_1$ the transition from $w=0$
to $w=-1$ occurs earlier, so that the models are 
favored from the data.

\begin{figure}
\includegraphics[height=2.9in,width=3.2in]{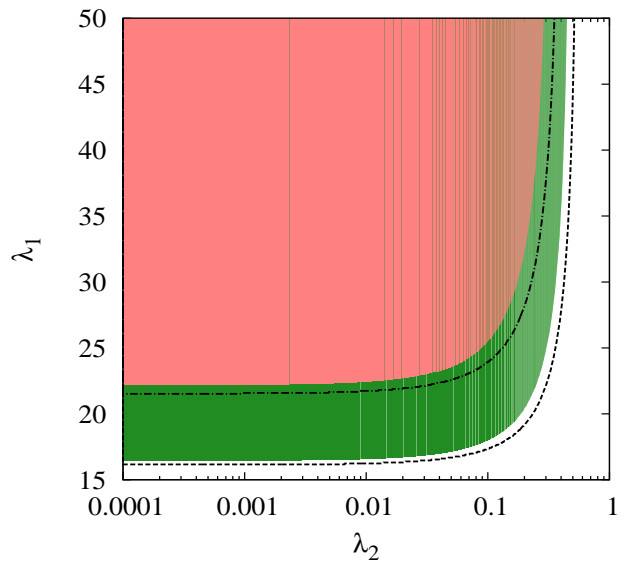}
\caption{\label{scalingcon3}
$1\sigma$ (red) and $2\sigma$ (green) observational contours
on $\lambda_1$ and $\lambda_2$ for the potential (\ref{doublepo}) 
with $\Omega_{m 0}=0.269$. 
The dot-dashed and dotted curves correspond to the $1\sigma$
and $2\sigma$ constraints without the BOSS data, respectively.}
\end{figure}

For non-zero values of $\lambda_2$ we do not have an
analytic expression of $w$, so we solve the 
background equations of motion numerically with
the priors $\lambda_1>9.4$ and $\lambda_2>10^{-4}$.
We set the latter prior because the $\lambda_2=0$ case
was already discussed above.
Varying the three parameters $\lambda_1$, $\lambda_2$, and
$\Omega_{m0}$, we find that the best-fit model parameters 
are $\lambda_1=54.94$, $\lambda_2=10^{-4}$, and 
$\Omega_{m0}=0.269$ with $\chi_{\rm min}^2=574.18$.
Then the Akaike criterion gives $\tilde{\chi}^2=574.18+2\times3=580.18$, 
which is larger than the $\Lambda$CDM value 
$\tilde{\chi}^2_{\Lambda{\rm CDM}}=575.89$
with the difference more than 4.
We also obtain the following bounds
\begin{eqnarray}
0.262&<&\Omega_{m0}<0.276\,\quad \textrm{(68\% CL),}\\
0.256&<&\Omega_{m0}<0.279\,\quad \textrm{(95\% CL),}\\
\lambda_1&>&16.3\,\quad \textrm{(68\% CL),}\\
\lambda_1&>&11.7\,\quad \textrm{(95\% CL),}\\
\lambda_2&<&0.361\,\quad \textrm{(68\% CL),}\\
\lambda_2&<&0.539\,\quad \textrm{(95\% CL).}
\end{eqnarray}

Figure \ref{scalingcon3} shows the observational constraints 
in the $(\lambda_1, \lambda_2)$ plane for $\Omega_{m0}=0.269$.
If the three parameters $\lambda_1$, $\lambda_2$, 
$\Omega_{m0}$ are varied in the likelihood analysis, 
it is difficult to marginalize over $\Omega_{m0}$
in the range $0<\Omega_{m0}<1$
because the solutions are prone to numerical instabilities 
around the tail regions of $\Omega_{m0}$.
Hence we use the fixed density parameter 
$\Omega_{m0}=0.269$, which corresponds to 
the best-fit value when the three parameters are varied.

If $\lambda_2 \gtrsim 0.5$, the model is excluded at 95 \% CL.
This is associated with the fact that $w$ possesses a minimum
for larger $\lambda_2$ (see Fig.~\ref{wscaling2}). 
Therefore, in the context of this model, the expansion of 
the Universe accelerates forever.
If $\lambda_1=20$, for example, the parameter $\lambda_2$
is constrained to be $\lambda_2 \lesssim 0.3$ (95 \%\,CL).
In fact the minimum of $w$ appears for $\lambda_2 \gtrsim 0.3$, 
which leads to the deviation of $w$ from $-1$ today.
In summary, the model is within the $2\sigma$
observational contour provided that 
$\lambda_2 \lesssim 0.1$ and $\lambda_1 \gg 1$. 

%%%%%%%%%%%%%%%%%%%%%%%
\begin{figure}[ht]
\includegraphics[height=2.9in,width=3.2in]{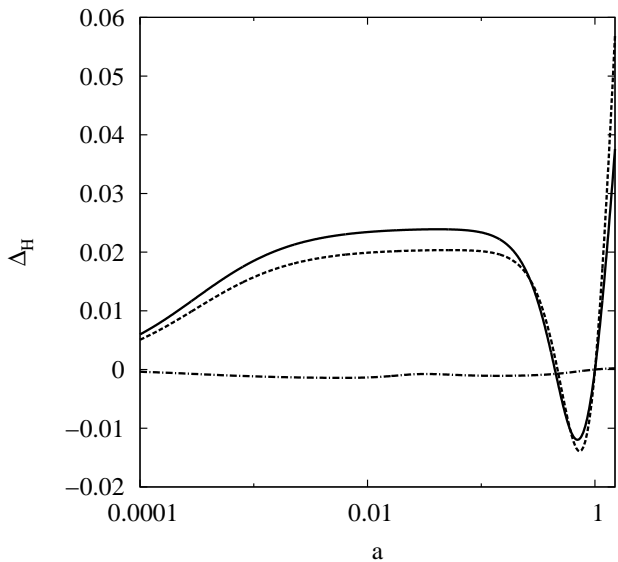}
\caption{\label{plotH}
Relative deviation of the Hubble parameter, 
$\Delta_H \equiv (H-H_{\Lambda {\rm CDM}})/H_{\Lambda {\rm CDM}}$,
for the best-fit cases of: (i) scaling (dot-dashed line), 
(ii) thawing (dotted line), (iii) tracker (solid line) models.
The relative deviations from the $\Lambda$CDM model are
less than 2.5\,\% for $a \leq 1$. 
The difference in the evolution among these models during 
dust domination, is due to the fact that different model parameters
lead to different values of $\Omega_{m0}$ for the best fit 
(in the $\Lambda$CDM we have $\Omega_{m0}=0.2699$, 
whereas the $\Omega_{m0}$'s for the other models exceed 
this value by about 4\,\%).}
\end{figure}
%%%%%%%%%%%%%%%%%%%%%%%

In Fig.~\ref{plotH} we plot the evolution of the relative deviation of the 
Hubble parameter from the $\Lambda$CDM as a function of $a$ 
for three best-fit models studied in this paper.
The deviation $\Delta_H$ of the best-fit scaling model is very small, which 
reflects the fact that $w$ quickly approaches $-1$ after the 
early transition from the scaling regime.
Meanwhile, in the best-fit thawing and tracker models, 
the relative deviations from the $\Lambda$CDM 
can reach the level of 2\,\%.

%%%%%%%%%%%%%%
\section{Conclusions}
\label{consec}
%%%%%%%%%%%%%%

In this paper we placed observational bounds on three types 
of quintessence models: (i) thawing, (ii) tracker, and (iii) scaling models.
We used the recent data of SN Ia from Union 2.1, the CMB shift 
parameters from WMAP7, 
and BAO from SDSS7 and BOSS, by which the background cosmic 
expansion history is constrained from the distance measurements. 

In thawing models where the field starts to evolve at the late cosmological 
epoch, the dark energy equation of state can be expressed as  
Eq.~(\ref{wtha}) with the three parameters $w_0$, $\Omega_{\phi0}$, 
and $K$. The parameter $K$ is related to the field mass at the 
initial stage. We put observational bounds in the $(w_0, \Omega_{\phi0})$
after marginalizing over $K$ (see Fig.~\ref{thawcon}).
If we take into account the BOSS data (\ref{bossdata}) in the likelihood
analysis, the upper bounds on $w_{0}$ are very close to $-1$ independent 
of the values of $K$ ranging in the region $0.1<K<10$ 
(under which the analytic formula (\ref{wtha}) is trustable).
Without the BOSS data the deviation of $w$ away from $-1$ today can 
be still allowed, as seen in Fig.~\ref{thawcon}.

In tracking freezing models where $w$ is nearly constant ($w \approx w_{(0)}$) 
during the matter era, there is the analytic formula (\ref{wtrack}) derived by 
considering a homogeneous perturbation around the tracker.
Without the BOSS data the tracker equation of state is constrained to be 
$-1.188<w_{(0)}<-0.949$ (95 \%\,CL), whose upper bound is tighter
than $w_{(0)}<-0.90$ derived by using SN Ia and SDSS7 data \cite{Chiba10}.
Inclusion of the BOSS data gives the upper bounds of $w_{(0)}$ close to $-1$.
We find that without the prior on $w_{(0)}$ the constraint is 
$-1.211<w_{(0)}<-0.998$ (95 \%\,CL) and with the quintessence
prior $w_{(0)}>-1$ the upper bound is $w_{(0)}<-0.964$ (95 \%\,CL).

For the potential (\ref{doublepo}) with $\lambda_1 \gg 1$ and 
$\lambda_2 \ll 1$, $w$ is close to 0 during the 
deep matter era because of the dominance of the steep 
potential $V_1 e^{-\lambda_1 \phi/M_{\rm pl}}$.
The field equation of state starts to decrease after the potential 
$V_2 e^{-\lambda_2 \phi/M_{\rm pl}}$ dominates over
$V_1 e^{-\lambda_1 \phi/M_{\rm pl}}$.
For larger $\lambda_1$ the exit from the scaling regime
($w=0$) occurs earlier. When $\lambda_2=0$ we found 
that the evolution of $w$ can be approximated by 
the parametrization (\ref{LHpara})
with $w_p=0$, $w_f=-1$, and $\tau \simeq 0.33$, where
the transition scale factor $a_t$ depends on $\lambda_1$.
Using this parametrization we derived the bound 
$a_t<0.23$ (95\,\%\,CL), which translates into 
the constraint $\lambda_1>13$.
This is consistent with the bound
shown in Fig.~\ref{scalingcon1} derived by solving the
background equations of motion numerically.

For the potential (\ref{doublepo}) with non-zero values of 
$\lambda_2$, the field equation of state tends to have a minimum for larger
$\lambda_2$. As we see in Fig.~\ref{scalingcon3}, the models
with $\lambda_2 \gtrsim 0.5$ is excluded at 95\,\%\,CL.
The parameters $\lambda_2 \lesssim 0.1$ and $\lambda_1>17$
are allowed from the data, which shows that the early transition 
from the scaling regime to the regime close to $w=-1$ is favored.
For the scaling models the observational constraints on $\lambda_1$
and $\lambda_2$ are not very sensitive to inclusion of the BOSS data, 
because $w$ evolves toward $-1$ at the late cosmological epoch.

As we see in Fig.~\ref{plotH}, the difference of $H$ between the 
three best-fit quintessence models and the $\Lambda$CDM 
model is only less than 2.5 \% in the past.
In current observations there is no strong evidence that 
quintessence is favored over the $\Lambda$CDM 
from the statistical point of view.
This property is especially significant by including the 
BOSS BAO data.

If we extend the analysis in the regime $w<-1$, we showed that 
some of the models studied in this paper can compete with the 
$\Lambda$CDM model according to the Akaike information criterion. 
We note that in many modified gravity models such as $f(R)$ gravity \cite{fR}, 
(extended) Galileons \cite{DT10}, and Lorentz violating theories \cite{Lorentz}
it is possible to realize $w<-1$ without having ghosts and instabilities.
It remains to see how much extent future high-precision observations
can constrain dark energy models
from the background expansion history as well as from the cosmic 
growth of density perturbations.

%%%%%%%%%%%%%%%%%%%%%%%%%%%%%%%%%%%%%
\section*{ACKNOWLEDGEMENTS}
AD thanks JSPS for financial support to visit Tokyo University 
of Science (No.~S12135). 
ST thanks warm hospitalities during his stays in 
Weihai, Observatorio Nacional in Rio de Janeiro, 
Passa Quatro, Szczecin, and University of Heidelberg. 
This work was supported in part by the Grant-in-Aid for
Scientific Research Fund of the 
JSPS No.~24540287 (TC) and No.~24540286 (ST) 
and Scientific Research on Innovative Areas (No.~21111006) (ST) 
and in part by Nihon University (TC). 
%%%%%%%%%%%%%%%%%%%%%%%%%%%%%%%%%%%%%

\end{document}